\documentclass[a4paper,twoside]{article}

\usepackage{epsfig}
\usepackage{subfigure}
\usepackage{calc}
\usepackage{amssymb}
\usepackage{amstext}
\usepackage{amsmath}
\usepackage{amsthm}
\usepackage{multicol}
\usepackage{pslatex}
\usepackage{apalike}
\usepackage{booktabs}
\usepackage{SCITEPRESS}     % Please add other packages that you may need BEFORE the SCITEPRESS.sty package.

\begin{document}

\title{The role of robot design in decoding error-related information from EEG signals of a human observer}

\author{\authorname{Joos Behncke\sup{1,2}, Robin T. Schirrmeister\sup{2}, Wolfram Burgard\sup{1} and Tonio Ball\sup{2}}
\affiliation{\sup{1}Department of Computer Science, Albert-Ludwigs-University Freiburg, Germany}
\affiliation{\sup{2}Translational Neurotechnology Lab, University Medical Center Freiburg, Germany}
\email{\{joos.behncke, robin.schirrmeister, tonio.ball\}@uniklinik-freiburg.de, burgard@informatik.uni-freiburg.de}
}

\keywords{EEG, Error processing, Deep Learning, Convolutional Neural Networks, Robot Design, Mirror Neuron System, BCI}

\abstract{For utilization of robotic assistive devices in everyday life, means for detection and processing of erroneous robot actions are a focal aspect in the development of collaborative systems, especially when controlled via brain signals. Though, the variety of possible scenarios and the diversity of used robotic systems pose a challenge for error decoding from recordings of brain signals such as via EEG. For example, it is unclear whether humanoid appearances of robotic assistants have an influence on the performance. In this paper, we designed a study in which two different robots executed the same task both in an erroneous and a correct manner. We find error-related EEG signals of human observers indicating that the performance of the error decoding was independent of robot design. However, we can show that it was possible to identify which robot performed the instructed task by means of the EEG signals. In this case, deep convolutional neural networks (deep ConvNets) could reach significantly higher accuracies than both regularized Linear Discriminanat Analysis (rLDA) and filter bank common spatial patterns (FB-CSP) combined with rLDA. Our findings indicate that decoding information about robot action success from the EEG, particularly when using deep neural networks, may be an applicable approach for a broad range of robot designs.}

\onecolumn \maketitle \normalsize \vfill

\section{\uppercase{Introduction}}
\label{sec:introduction}

\noindent Error-related brain activity has the potential to support the management and implementation of BCI systems in various scenarios. Error detection systems have been embedded to improve real life tasks, e.g., to detect and correct erroneous actions performed by an assistive robot online, using primary and secondary error-related potentials (ErrPs) \cite{salazar2017correcting}. Other approaches benefit from ErrPs to extract incorrect maneuvers in a real-world driving task \cite{zhang2015eeg} or use error-related negativity (ERN) to correct erroneous actions in object moving tasks \cite{parra2003response}. In addition, error-related brain signals have been utilized to teach neuroprosthetics appropriate behaviour in situations with changing complexity \cite{iturrate2015teaching}. 

Recently we could show that deep Convolutional Neural Networks (deep ConvNets) significantly improved the decoding of robot errors from the EEG of a human observer \cite{behncke2018signature}. However, the role of robot design in the decodability of error-related information in this scenario has not yet been investigated, although robots come in a broad range of different designs. Beside specifying properties like mobility or autonomy, a robot can be classified regarding its grade of being humanoid. Especially when collaborative interactions with humans take place, this characteristic may have an influence on the human's perception of the robots behaviour. In a comparison between a human and a robotic agent performing several movement tasks, an activation of the human mirror neuron system \cite{rizzolatti1996premotor} could be shown \cite{gazzola2007anthropomorphic}.

To investigate the impact of robot design, we designed a study where two robots performed exactly the same task, either correctly or not. The two robots differed in their physical appearance and basic design. We analyzed the classification performances of the error-related EEG signals separately for the different robot types. Furthermore, we investigated whether a distinction between the two robot types could be made based on the EEG recordings. For classification, we employed a deep ConvNet, and we compared decoding performances with those of two more conventional methods, namely a regularized Linear Discriminant Analysis (rLDA), see \cite{friedman1989regularized}, and filter bank common spatial patterns (FB-CSP) combined with an rLDA classifier.

\section{\uppercase{System and Experimental Design}}

\begin{figure}[ht]
\centerline{\includegraphics[width=0.5\textwidth]{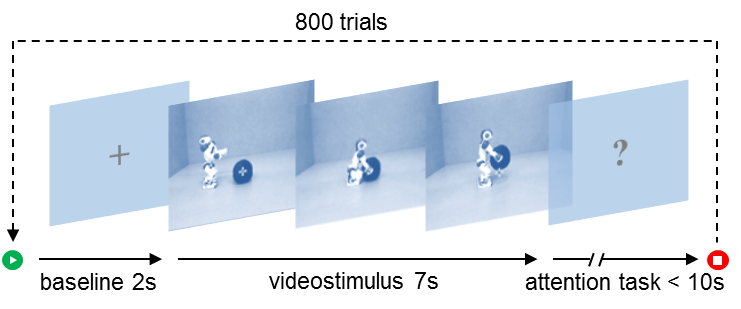}}
\caption{Timing structures of the experiment ($>800$ trials per participant). Each trial consisted of a $2\,s$ fixation period, video stimulus of $7\,s$ and an attention control task.}
\label{figpara}
\end{figure}

\noindent In this study, participants were instructed to observe two robots performing a naturalistic task either in a correct or an erroneous manner. A humanoid (NAO (TM), Aldebaran Robotics, Paris, France) and non-humanoid robot (referred to as NoHu) were programmed to approach, grab, and lift a ball from the ground. They either succeeded in lifting or failed, letting the ball drop again. The visual stimuli consisted of short video clips, which were presented to the participants in a randomized order. Each trial consisted of a fixation period followed by the video playback and completed by the final attention task. Figure \ref{figpara} shows the trials structure and the experimental paradigm schematically. A total of at least 800 trials per participant were recorded, split into sessions of 40 trials.

12 healthy participants (22-31 years old, 6 female) participated in our experiments, giving their informed consent for the study, which was approved by the local ethics committee. While observing the performed tasks, the participants were seated inside a dimly lit EEG recording chamber, being equipped with an active electro-magnetic shielding to ensure a high signal quality. The EEG cap consisted of 128 gel-filled electrodes positioned according to the five-percent electrode layout. 

\section{\uppercase{Methods}}

\subsection{Pre-processing and Decoding}
\label{preproc}

\noindent The recorded EEG signals were re-referenced to a common average reference (CAR) and resampled to $250\,Hz$. An electrode-wise exponential moving standardization was applied to normalize the data by exponential moving means and variances. Then, according to stimulus onset and predefined decoding intervals the data were cut into trials and used as classification features. The deep ConvNets were designed using the open-source deep learning toolbox \textit{braindecode} for raw time-domain EEG, based on the model \textit{Deep4Net}. In the first of the four convolutional-max-pooling blocks a temporal and a spatial convolution over all channels is done stepwise. Followed by three conventional convolution blocks, the output is finally delivered by a dense softmax classification layer. More detailed information about the underlying architecture can be found in \cite{schirrmeister2017deep}. The data were split into two sets, $80\,\%$ were used for training while the remaining $20\,\%$ served as a test set.

For the rLDA the pre-processing was similar except the fact that no standardization was implemented, resulting in higher accuracies than with standardization. The rLDA algorithm comprised a shrinkage parameter estimation, based on the \textit{LedoitWolf} estimator \cite{ledoit2004well} as suggested in \cite{blankertz2010berlin}

In the FB-CSP implementation, the data were high-pass filtered using a \textit{Butterworth} filter (order $4$, cut-off frequency $0.5\,Hz$) and automatically cleaned before cut into trials. The resulting trial structure was then band-pass filtered in $35$ non-overlaying frequency bands between $0.5\,Hz$ and $144\,Hz$, having a bandwidth of $2\,Hz$ for bands lower than $30\,Hz$ and a bandwidth of $6\,Hz$ for higher frequency bands \cite{buzsaki2004neuronal}. For further details on feature selection and classifier design, see \cite{ang2012filter,blankertz2008optimizing}.

\subsection{Decoding Intervals}
\label{decint}

Relative to the video onset, two decoding intervals were selected. A long decoding interval of $0-7\,s$ covered the full length of the video, while a late interval of $4-7\,s$ was selected since it was covering the actual process of grasping and lifting the ball. For both intervals decoding performances were calculated using all of the methods presented in \ref{preproc}.

\subsection{Statistical Procedures}

For estimation of significance of the decoding results per participant a random permutation test was applied \cite{pitman1937significance}. Each of the $n=10^6$ sets of randomly shuffled labels, reflecting the real total number of trials per class, was compared to the genuine labels to create a reasonable distribution of potential classification results. The generated distribution was then used to determine the significance per participant and decoding interval. In Figure \ref{figallacc} group significances were estimated by means of the single trials and using a sign test \cite{hollander2013nonparametric}.

\section{\uppercase{Error Decoding}}

\begin{figure}[ht]
\centerline{\includegraphics[width=0.5\textwidth]{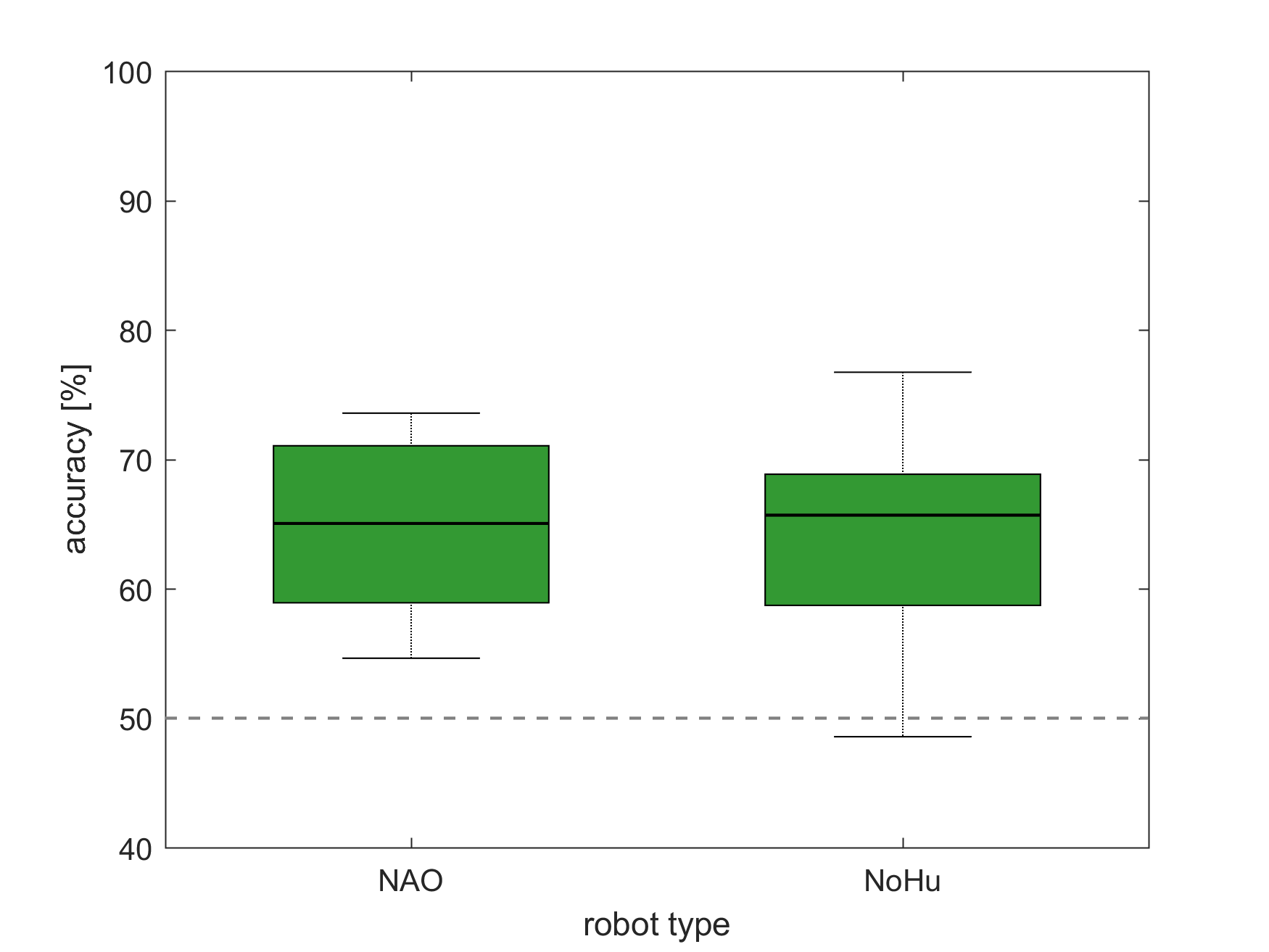}}
\caption{Boxplots of accuracies for error decoding for the two robot types.}
\label{figallacc}
\end{figure}

Firstly, we investigated the influence of robot design on the performance in an error-decoding scenario. Therefore, the trials were sorted by robot type and the decoding analysis of erroneous vs. correct trials was performed separately on the two data sets. In this case, only the deep ConvNet implementation was used for error detection. Figure \ref{figallacc} shows the outcome of this analysis, showing the distribution of the accuracies for both robot conditions for all participants. Even though the decoding of trials presenting the NoHu robot executing the task showed a broader range, there was no significant ($p\,=\,1$) difference between the two conditions. The ConvNets achieved median accuracies of $(64.82\,\pm\,6.75)\,\%$ for the NAO condition and $(64.02\,\pm\,8.72)\,\%$ for the NoHu condition, yielding almost the same performance taking the errors into account.

\section{\uppercase{Distinction between robot types}}

\noindent Three different decoding methods were used for classification, ConvNets, rLDA and FB-CSP+rLDA. For each participant decoding accuracies were determined for both decoding intervals (see \ref{decint}) which were defined according to the video onset. An overview of mean decoding accuracies is given in Table \ref{tabcomp}. The results indicate that a distinction between the two kinds of robots was decodable with all of the applied techniques. Furthermore, for both decoding intervals the deep ConvNet architecture performed consistently and significantly better (sign test, $p\,<\,0.01$). 

\begin{table}[hb]
\caption{Mean Accuracies for Different Decoding Intervals}
\begin{center}
\begin{tabular}{cccc}
\toprule
\textbf{}&\textbf{0 - 7s}&\textbf{4 - 7s}\\
\midrule
\textbf{ConvNet} & $(78.31\,\pm\,8.09)\,\%$ & $(73.80\,\pm\,7.52)\,\%$ \\ 
\textbf{rLDA} & $(68.29\,\pm\,8.00)\,\%$ & $(64.71\,\pm\,7.36)\,\%$ \\ 
\textbf{FB-CSP} & $(55.71\,\pm\,4.54)\,\%$ & $(56.80\,\pm\,3.92)\,\%$ \\ 
\bottomrule
\end{tabular}
\label{tabcomp}
\end{center}
\end{table}

\begin{figure}[ht]
\centerline{\includegraphics[width=0.5\textwidth]{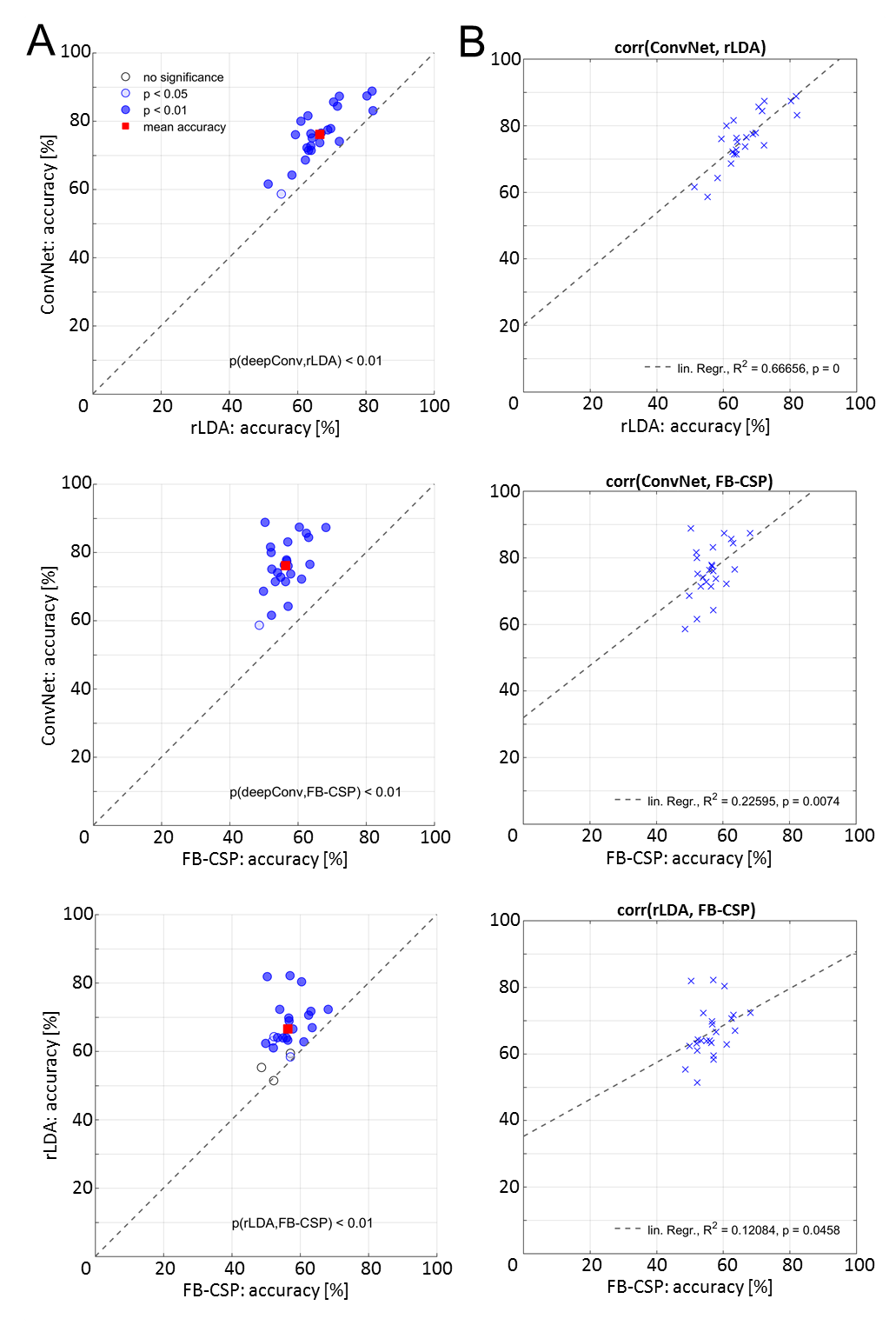}}
\caption{(A) Pairwise comparison of decoding accuracies investigated with three different methods. (B) Pairwise linear regression of the participants performances.}
\label{figrobacc}
\end{figure}

Figure \ref{figrobacc}A shows the pairwise comparison of performances for the different classifiers, including the results of the analysis gained with both of the decoding intervals. Significance is indicated by blue color while the red squares represent the mean accuracies, the diagonal indicates equal performance. For each single participant, the decoding accuracies gained by the deep ConvNet implementation significantly exceeded those of the other methods, i.e., compared to rLDA and FB-CSP, the ConvNet performed significantly better on the group level. 

To gain a measure of the relationship between the decoding results of the different decoding techniques, we calculated the pairwise linear correlation between decoding accuracies over subjects. Figure \ref{figrobacc}B shows the correlation for all pairs; the dotted line represents the result of the linear regression. The results indicate a highly significant linear relationship between deep ConvNet and rLDA performances, but also for both of the other combinations a significant ($p\,<\,0.05$) correlation can be found. 

\section{\uppercase{Visualization}}

\begin{figure*}[ht]
\centerline{\includegraphics[width=\textwidth]{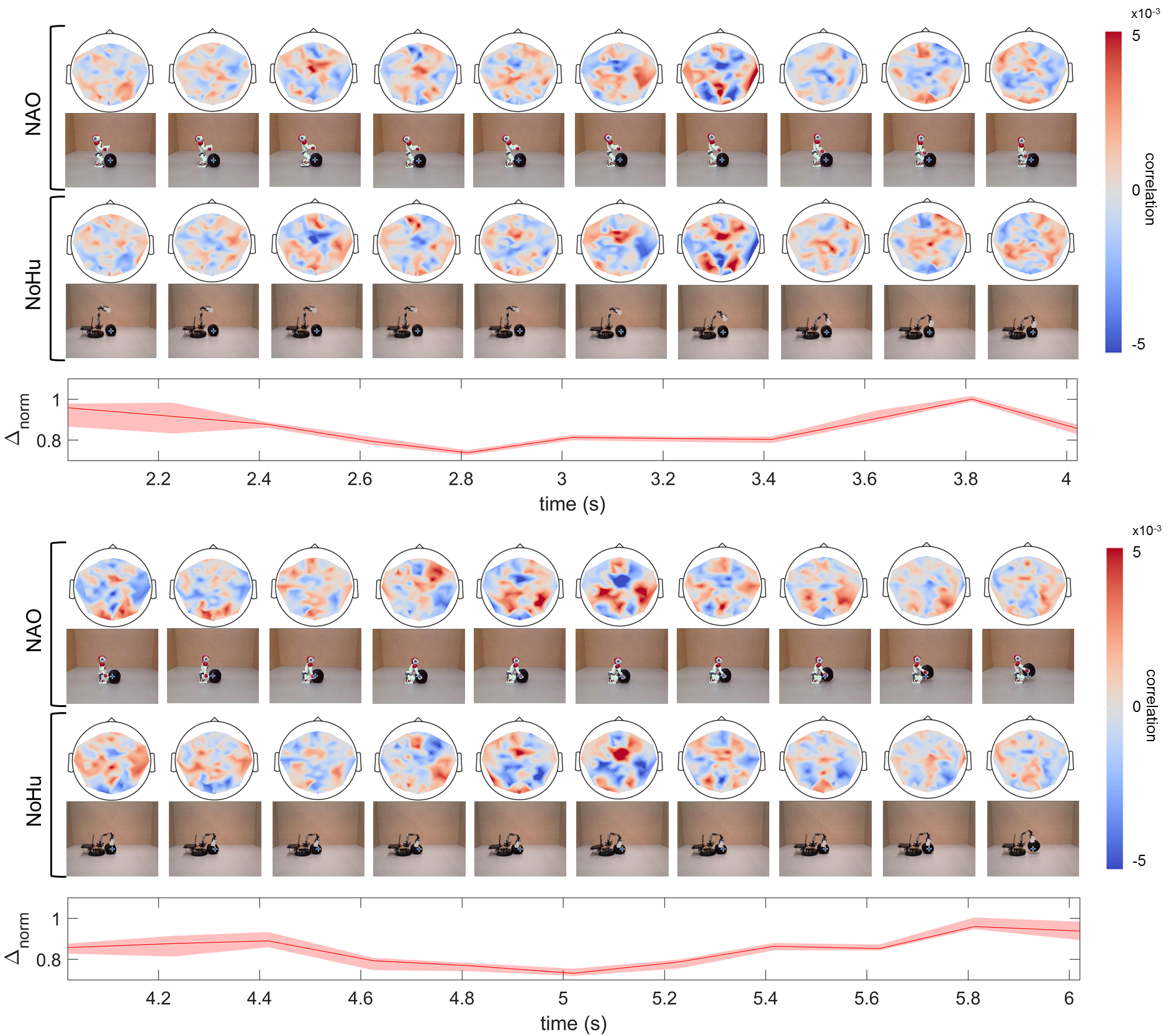}}
\caption{Time-resolved voltage feature input-perturbation network-prediction correlation maps for robot type decoding, averaged over participants and 30 iterations, and corresponding video frames (top rows). Time-resolved normalized L1 distance of sequential pairs of video frames for both conditions (bottom row). }
\label{figpertmaps}
\end{figure*}

The visualization used in this section was applied as suggested in \cite{schirrmeister2017deep}, visualizing EEG features that the deep ConvNet learned from the data when used to distinguish between the different robot designs. To this aim, the correlation between changes of the predictions made by the ConvNet and perturbation changes in time domain voltage amplitudes were calculated. By adding Gaussian noise, training trials were randomly perturbed. After that, the perturbed signal and the unperturbed signal were both fed into the deep ConvNet. The two outputs were extracted before the softmax activation, and their difference was correlated with the perturbation itself. The results can be visualized in a channel-resolved manner. The decision to use time domain signals as an input for the perturbations instead of spectral amplitudes was made according to the fact of lower decoding performances of the CSP algorithm, which relies on spectral features. This circumstances suggest that spectral information was likely not that prominent in the underlying decoding problem. The correlations shall indicate at which moment changes in certain channels might contribute to the classifiers decision causally.

In Figure \ref{figpertmaps} the (participant- and iteration-) averaged time-resolved input-perturbation network-prediction correlation maps for voltage features of the robot type decoding are shown. The maps are depicted from $2-6\,s$, whereby each map illustrates the correlation of a $0.2\,s$ time bin. For each bin, the two maps (NAO and NoHu) are shown together with their corresponding video frame. As expected, pairs of maps for the two different conditions exhibit opposed correlations. Figure \ref{figpertmaps} is divided into a section were the robots approach the ball (top) and a section were the robots try to lift the ball (bottom). At the bottom of each section, the time-resolved normalized L1 distance $\Delta_{norm}$ of sequential pairs of video frames for both conditions is illustrated.

The perturbation maps in Figure \ref{figpertmaps} exhibit increasing, prominent correlation patterns for signals around $3.2\,s$ and $5.0\,s$ according to video onset. At the first time point, the effects show spatially more widespread correlations with a remarkable, centered peak in frontal areas accompanied by an occipital symmetric effect. The later time window around $5.0\,s$ shows similar but less symmetric patterns in occipital regions, and a more pronounced frontal peak. The occipital effects for both time points might indicate different cerebral processings of the visual characteristics in the robots execution of the programmed task.

As a first step, to examine visual features which might have led to differences in brain signals for the two robot conditions, we calculated the time-resolved normalized L1 distance $\Delta_{norm}$. The corresponding curve in Figure \ref{figpertmaps} indicates a rather small difference between sequential frames and a steady difference between the two conditions, as the curve varies only little but with consistently high values. Furthermore, with this method no obvious correlation between the time course of visual changes and informative features extracted by the perturbation analysis can be suggested. 

\section{\uppercase{Conclusion \& Outlook}}
\label{sec:conclusion}

\noindent The potential of error detection for practical BCI applications has recently led to several studies in EEG research, e.g. \cite{salazar2017correcting,zhang2015eeg,parra2003response,iturrate2015teaching}. However, to our knowledge, aspects of robot design has not been part of investigations on error-related brain signals so far. In this study, we examined this aspect and show that at least the two different robot designs used in our current study did not have any significant negative impact on the performance of error decoding, indicating that in principle error decoding is applicable to various robot types. 

The distinction between robot types based on the participants brain-signals yielded in mean accuracies of $(78.31\,\pm\,8.09)\,\%$ for the ConvNet, $(68.29\,\pm\,8.00)\,\%$ for the rLDA and $(55.71\,\pm\,4.54)\,\%$ for FB-CSP implementation when decoding on the whole time-window covered by the video stimulus. Hence, the robot type indeed could be classified on the basis of the EEG data. Moreover, our ConvNet implementation could reach systematically better results than the two other methods. Notably, the ConvNet was not specifically adjusted to the problem, but used "out-of-the-box" as it was available in the used open-source implementation. Deep ConvNets have already entered the field of EEG research and proven their applicability \cite{tang2017single}, but not yet for distinction of robot types. Overall, the accuracies reached here are however still far from the requirements for practical application. E.g. intracranial signals or further improved non-invasive methods will be necessary to reach better performance. 

The pairwise comparison of the  participants decoding results for the different methods showed a significant linear correlation for all cases. Particularly the behaviour of ConvNet und rLDA accuracies were apparent: the net appears to act more "rLDA-like" for the given problem, in line with the assumption that time-domain features were mainly informative.

For the visualization, the time-domain EEG signal changes were correlated with decoding results determined after a Gaussian perturbation. Based on the correlation maps as shown in Figure \ref{figpertmaps} it appears that the ConvNet learns time-domain information to distinguish between classes, as specific time windows reveal more pronounced effects than others. Thus, a prominent focus of the distribution of informative signals lies on occipital brain regions. This might reflect the difference in the processing of the visual input for the two robot types. Visualizations also show a distinct medio-frontal peak. This effect appeared around the same time when the robot approached to the target and grasped it. It is possible that in this cases the mirror neuron system (MNS) might get activated \cite{hari1998activation,rizzolatti1996premotor,rizzolatti2004mirror,mukamel2010single}, which involves widespread frontal and parietal regions. According to this, the different robot types could have lead to a differential activation of this system.

%\vfill
\section*{\uppercase{Acknowledgements}}

\noindent This work was supported by DFG grant EXC1086 BrainLinks-BrainTools and Baden-Württemberg Stiftung grant BMI-Bot. Furthermore, the authors would like to thank Dominik Welke and Marina Hader for their support in creating and processing the data sets, but especially the subjects for their conscientious participation. 
%\vfill
\bibliographystyle{apalike}
{\small
\bibliography{main}}

%\section*{\uppercase{Appendix}}

%\noindent If any, the appendix should appear directly after the
%references without numbering, and not on a new page. To do so please use the following command:
%\textit{$\backslash$section*\{APPENDIX\}}

\vfill
\end{document}